\documentclass[runningheads]{llncs}
\usepackage{numprint}
\usepackage{siunitx,etoolbox}
\usepackage{comment}
\usepackage{ifthen}
\usepackage{multirow}
\npthousandsep{,}
\nprounddigits{2}
\usepackage{graphicx}
\usepackage{booktabs}
\usepackage{siunitx}
\usepackage{etoolbox}% <-- new
\newcommand{\ubold}{\fontseries{b}\selectfont} % renew def. for non-extended bold font
\robustify\ubold

\makeatother

\newcommand\blfootnote[1]{%
  \begingroup
  \renewcommand\thefootnote{}\footnote{#1}%
  \addtocounter{footnote}{-1}%
  \endgroup
}

\begin{document}
\title{Generative adversarial network for segmentation of motion affected neonatal brain MRI }

\author{N. Khalili\inst{1} \and E. Turk\inst{2} \and M. Zreik \inst{1} \and M.A. Viergever \inst{1,3} M.J.N.L. Benders\inst{2,3} \and I. I\v sgum \inst{1,3}}

\institute{Image Sciences Institute, University Medical Center Utrecht, The Netherlands\\\and Department of Neonatology, Wilhelmina Children’s Hospital, University Medical Center Utrecht, The Netherlands\\\and Brain Center Rudolf Magnus, University Medical Center Utrecht, The Netherlands\\}
\authorrunning{Khalili et al.}% Part of LEFT running header
\titlerunning{GAN for segmentation of motion affected neonatal brain MRI}% Part of RIGHT running header
\maketitle              % typeset the header of the contribution
\begin{abstract}
\blfootnote{\small\textbf{Accepted in Medical Image Computing and Computer Assisted Intervention 2019}}
Automatic neonatal brain tissue segmentation in preterm born infants is a prerequisite for evaluation of brain development. However, automatic segmentation is often hampered by motion artifacts caused by infant head movements during image acquisition. Methods have been developed to remove or minimize these artifacts during image reconstruction using frequency domain data. However, frequency domain data might not always be available. Hence, in this study we propose a method for removing motion artifacts from the already reconstructed MR scans. The method employs a generative adversarial network trained with a cycle consistency loss to transform slices affected by motion into slices without motion artifacts, and vice versa. In the experiments 40 T2-weighted coronal MR scans of preterm born infants imaged at 30 weeks postmenstrual age were used. All images contained slices affected by motion artifacts hampering automatic tissue segmentation. To evaluate whether correction allows more accurate image segmentation, the images were segmented into 8 tissue classes: cerebellum, myelinated white matter, basal ganglia and thalami, ventricular cerebrospinal fluid, white matter, brain stem, cortical gray matter, and extracerebral cerebrospinal fluid. Images corrected for motion and corresponding segmentations were qualitatively evaluated using 5-point Likert scale. Before the correction of motion artifacts, median image quality and quality of corresponding automatic segmentations were assigned grade 2 (poor) and 3 (moderate), respectively. After correction of motion artifacts, both improved to grades 3 and 4, respectively. The results indicate that correction of motion artifacts in the image space using the proposed approach allows accurate segmentation of brain tissue classes in slices affected by motion artifacts. 

\keywords{motion correction \and convolutional neural network \and cycleGAN \and neonatal MRI}
\end{abstract}
\section{Introduction}
\begin{figure}[t]
\includegraphics[width=\textwidth]{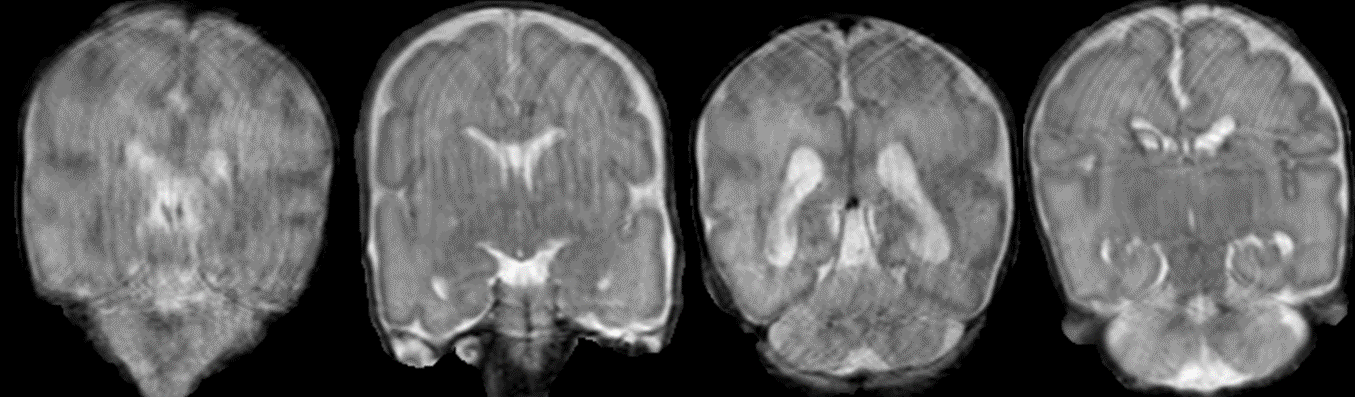}
\caption{Examples of coronal slices from T2-weighted MRI acquired in preterm born infants at 30 weeks postmenstrual age affected by motion artifacts. Structures outside the neonatal cranium have been masked out.}
    \label{fig:example}
\end{figure}

Important brain development occurs in the last trimester of pregnancy including brain growth, myelination, and cortical gyrification \cite{kostovic2006development}. Magnetic resonance imaging (MRI) is widely used to non-invasively assess and monitor brain development in preterm infants. In spite of ability of MRI to visualize the neonatal brain, motion artifacts caused by the head movement lead to blurry image slices or slices with stripes (see Figure \ref{fig:example}). These artifacts hamper image interpretation as well as brain tissue segmentation.

To enable the analysis of images affected by motion artifacts, most studies perform the correction in the frequency domain (k-space) prior to analysis \cite{atkinson1997automatic,godenschweger2016motion}. However, frequency domain data is typically not stored and hence, not available after image reconstruction. Recently, Duffy at al. \cite{duffy2018retrospective} and Paware et al. \cite{pawar2018moconet} proposed to use convolutional neural networks (CNNs) to correct motion-corrupted MRI from already reconstructed scans. CNNs were trained to reconstruct simulated motion artifacts that were modelled with a predefined formula. This enforces the network towards an assumed distribution of artifacts. However, in practice, it is difficult to estimate the real distribution of motion. Alternatively, a CNN could be trained to generate images without motion artifacts from images with such artifacts. However, this would require training with paired scans, which are rarely available. To solve this, recently cycleGAN has been proposed to train CNNs for image-to-image transformation with unpaired images \cite{zhu2017unpaired}. %These CNNs have been shown to successfully transform images from one domain to another domain in natural images \cite{zhu2017unpaired}, as well as in medical images \cite{wolterink2017deep,chartsias2017adversarial}.

In this study, we propose to employ a cycleGAN to generate MR slices without motion artifacts from slices affected by motion artifacts in a set of neonatal brain MR scans. The cycleGAN is trained to transform slices affected by motion artifacts into slices without artifacts, and vice versa. To generate slices corrected for motion artifacts, we applied the trained cycleGAN to motion affected slices and we hypothesize that images corrected for motion artifacts allow more accurate (automatic) segmentation. To evaluate this, we use a method exploiting a convolutional neural network to segment scans into eight tissue classes. Moreover, we propose to augment the segmentation training data from the cycleGAN that synthesizes slices with artifacts from slices without the artifacts. We demonstrate that the proposed correction for motion artifacts improves image quality and allows accurate automatic segmentation of brain tissue classes in brain MRI of infants. We also show that the proposed data augmentation further improves segmentation results.

%segment the resulting scans into 8 tissue classes using a CNN. Additionally, the segmentation network was trained with motion augmented data using the motion-generator cycle. The motion generator cycle randomly synthesized motion on motion-free images to robust segmentation performance against motion artifacts.

%Additionally, the segmentation network was trained with motion augmented data using the motion generation network to make segmentation performance robust toward motion artifacts.

\section{Data}
This study includes 80 T2-weighted MRI scans of preterm born infants scanned at average of $30.7\pm1.0$ weeks postmenstrual age (PMA). Images were acquired on a Philips Achieva 3T scanner at University Medical Center Utrecht, the Netherlands. The acquired voxel size was $0.34\times0.34$ mm$^2$ and the reconstruction matrix was $384\times384\times50$. The scans were acquired in the coronal plane. In this data set, 60 scans had visible motion artifacts in most of the slices and 20 scans had no visible motion in any slice. The reference segmentation of 10 scans out of 20 scans without motion artifacts were available. The scans were manually segmented into 8 tissue classes: cerebellum (CB), myelinated white matter (mWM), basal ganglia and thalami (BGT), ventricular cerebrospinal fluid (vCSF), white matter (uWM), brain stem (BS), cortical gray matter (cGM), and extracerebral cerebrospinal fluid (eCSF).% - following the neonatal brain segmentation protocol described in the NeoBraiS12 challenge \cite{ivsgum2015evaluation}. 

\section{Method}
Motion artifacts in the neonatal brain MR hamper the diagnostic interpretability and precise automatic segmentation of the brain tissue classes. To address this, we propose to correct motion artifacts in the reconstructed MR scans using a cycleGAN. Thereafter, to evaluate whether the corrected images are suitable for segmentation of brain tissues, a CNN architecture was trained to segment the brain into eight tissue classes. Furthermore, to improve segmentation performance, we proposed to augment the training data by synthesizing images with motion artifacts from the images without artifacts using the cycleGAN.

\subsection{Artifact correction network}

\begin{figure}[t]
\centering
\includegraphics[width=\textwidth]{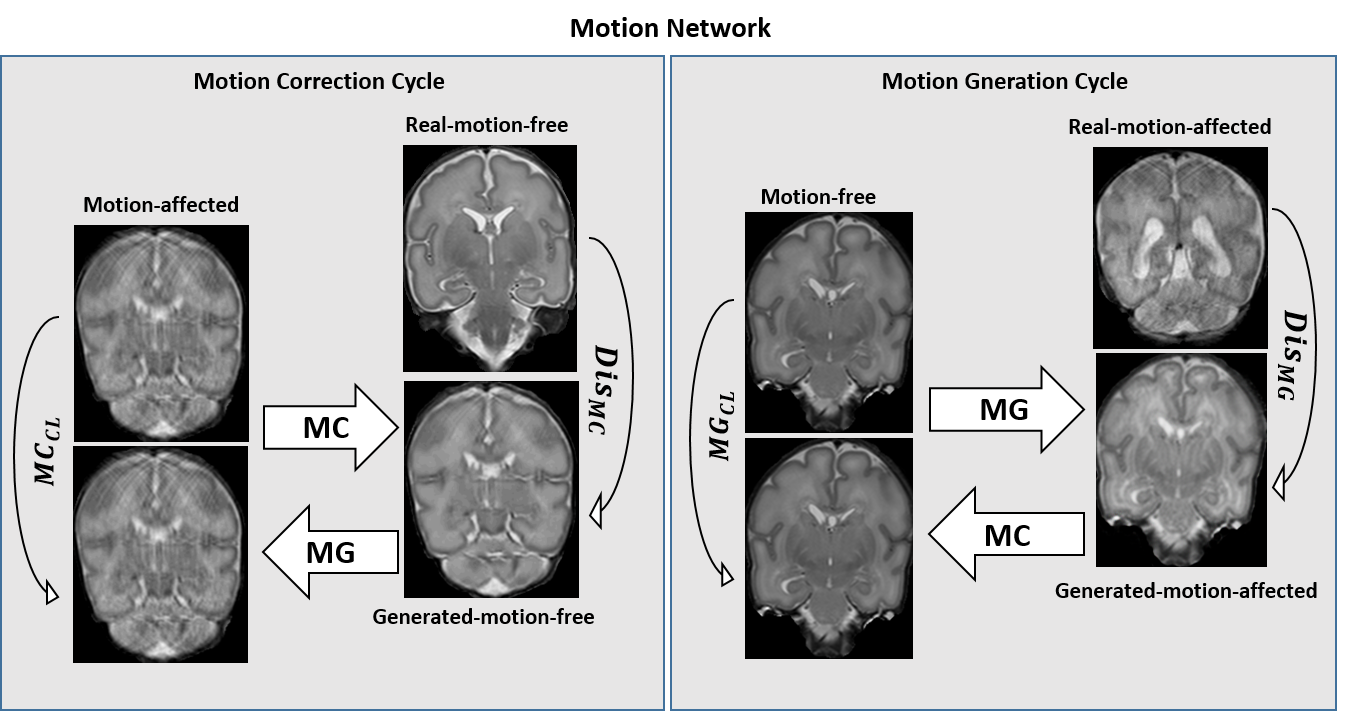}
\caption{ The CycleGAN consists of two cycles: motion correction and motion generation. In the motion correction cycle, first network is trained to transform slices affected by motion into slices without motion artifacts ($MC$), the second network is trained to transform the generated slices without motion artifacts back to the original slices ($MG$), and the third network discriminates between real and synthesized slices without motion artifacts ($Dis_{MC}$). In the motion generation network, motion was added to the slices without motion artifacts ($MG$), motion correction network transforms generated slices to the original slices ($MC$), and the discriminator network discriminates between real and fake slices affected by motion artifacts ($Dis_{MG}$)}
    \label{fig:cycleGAN}
\end{figure}

CycleGAN has been proposed to train image-to-image translation CNNs with unpaired images. Given that obtaining paired scans with and without motion artifacts is difficult, cycleGAN was trained to transform slices affected by motion to slices without motion artifacts and, vice versa (Figure \ref{fig:cycleGAN}). The network architecture consists of two cycles, motion correction and motion generation cycles. The motion correction cycle consists of three networks. Motion correction network ($MC$) transforms slices affected by motion to slices without motion artifacts. Motion generation network ($MG$) reconstructs the generated slices without motion artifacts to the original image slices. A discriminator CNN discriminates between generated and real slices without motion artifacts $Dis_{MC}$. While the discriminator distinguishes between generated and real slices without motion artifacts, the generator tries to prevent it by generating images which are not distinguishable for the discriminator. Similarly, motion generation cycle transforms slices without motion artifacts to slices affected by motion. The network architecture in both cycles is identical. The generator contains 2 convolution layers with stride of 2, 9 residual blocks \cite{he2016deep}, and 2 fractionally strided convolutions with stride proposed in \cite{johnson2016perceptual}. The discriminator networks have a PatchGAN \cite{isola2017image}, which classifies $70 \times 70$ overlapping image patches as fake or real. Two adversial losses \cite{goodfellow2014generative} were used in both motion correction network and motion generation network. Furthermore, cycle consistency loss in motion correction network ($MC_{cl}$) and motion generation network ($MG_{cl}$) were weighted by $\lambda$ and were added to adversial losses.

%The discriminator $D_{MC}$ aim to predict the label 1 for real motion free slices and the label 0 for corrected slices. Therefore, the discriminator loss function of motion correction cycle $Corr$ is
%\begin{equation}
 %   L_{MC}= (1- D_{MC}(I_{MF})))^2 + D_{MC}(Corr(I_{M}))^2
%\end{equation}

%where $I_{MF}$ is motion free slices and $I_{M}$ is slices with motion. $Corr$ is mapping function that transfer motion-affected slices to motion-free or in another word is a motion correction function.

%A similar adverial loss was introduced for motion generation cycle $Gen$ where $D_{MG}$ aim to predict the label 1 for slices with real motion and 0 for generated motion slices. 
%\begin{equation}
%    L_{MG}= (1- D_{MG}(I_{M})))^2 + D_{MG}(Gen(I_{MF}))^2
%\end{equation}https://www.overleaf.com/project/5c6d0a83effbc01508a784cb
 
%To enforce cycle consistency between two motion and motion free domains, the consistency loss consists of forward and backward cycle consistency during the training. 
%\begin{equation}
%    L_{cycle}= ||Gen(Corr(I_M))-I_M||_1 + ||Corr(Gen(I_{MF}))-I_{MF}||_1
%\end{equation}

%The total loss is summation of adversial losses and cycle loss weighted by parameter $\lambda$ . 

\subsection{Segmentation Network}
To assess segmentation performance in images affected by motion artifacts, a CNN with Unet-like architecture was trained to segment images into eight tissue classes. The segmentation network consists of a contracting path and an expanding path. The contracting path consists of 10 $3\times3$ convolution layers followed by rectified linear units (ReLUs). Every two convolution layers the features were downsampled by $2\times2$ max pooling and the feature channels were doubled using the following scheme 32, 64, 128, 256, 512. In the expanding path, an up-sampling is followed by a $2\times2$ convolution which halves the number of feature channels. The results are concatenated with the corresponding contracting path and convolved by two $3\times3$ convolutional layers followed by a ReLU. In the final layer, one $1\times1$ convolutional layer maps each component of the feature vector to the desired number of classes. Batch normalization is applied after all convolutional layers to allow for faster convergence. The network was trained with 3D patches of $256 \times 256 \times 3$ voxels. The network was trained by minimizing the average of Dice coefficient in all classes between the network output and manual segmentation.

\section{Evaluation}

 Given that slices affected by motion don't allow accurate manual annotation, to quantitatively evaluate the proposed method, motion is synthesized in images using the motion generation network. This allows evaluation with the manual annotations performed in images without artifacts. Thereafter, the performance of the segmentation network was evaluated using the Dice coefficient (DC), Hausdorf distance (HD) and mean surface distance (MSD) between manual reference and automatically obtained segmentations. The evaluation was performed in 3D. 
 
 To evaluate the proposed method on images with real motion artifacts, the images and the corresponding automatic segmentations before and after motion correction were qualitatively evaluated using 5-points Likert scale. The image quality was scored on a scale from 1 to 5, where 1 indicates uninterpretable images with severe motion artifacts, and 5 indicates excellent image quality. Similarly, automatic segmentations were scored 1 when the segmentation failed, and 5 when the segmentation was very accurate.

\section{Experiments and Results}
Prior to analysis, the intracranial brain volume was extracted from all scans using Brain Extraction Tool \cite{smith2002fast}. To train the artifact correction network, 15 scans without motion artifacts and 20 scans with motion artifacts were selected for training. The remaining 5 scans without motion artifacts and 40 scans with motion artifacts were used for testing. From scans without motion artifacts, 700 slices without visible artifacts were selected. Similarly, from the scans with motion artifacts, 714 slices with visible artifacts were selected. The network was trained with a batch size of 4. Adam \cite{kingma2014adam} was used to minimize the loss function for 100 epochs with a fixed learning rate of 0.00005. $\lambda$ was set to 10. 

To segment the brain into eight tissue classes, the segmentation network was trained with 5 scans without motion artifacts selected from the 15 training scans used to train the motion correction network. The segmentation network was trained with a batch size of 6. Adam was used to minimize the loss function for 200 epoch and the learning rate was set to 0.0001.

In the experiments, we performed quantitative evaluation of the proposed method through the evaluation of the brain tissue segmentation. First, to determine the upper limit of the segmentation performance, images without artifacts were segmented (Table \ref{tab:qantitative}, top row). Second, we aimed to evaluate the segmentation performance in the images with artifacts. However, motion artifacts are prohibitive for accurate manual annotation thus, those were not available for such images. Hence, the motion generation network was used to synthesize images with artifacts from the images without artifacts, for which manual segmentations were available. Segmentation was performed in the synthesized images. (Table \ref{tab:qantitative}, second row). Third, using motion correction network, the artifacts were removed from the images with synthesized artifacts and those were subsequently segmented (Table \ref{tab:qantitative}, third row). 
In the previous experiments, the segmentation network was trained only with images without motion artifacts, as only those were manually labelled. However, we hypothesized that the performance would improve when the segmentation would be trained with both types of images. Hence, to obtain images affected by motion that can be used for training, similar to the second experiment, we synthesized training images using motion generation network. In the fourth experiment, we evaluated segmentation network trained with augmented training data, i.e. images with and without motion artifacts on images with synthesized motion artifacts (Table \ref{tab:qantitative}, fourth row). Finally, segmentation was performed in images with corrected synthesized artifacts as in the third experiment, and training data for the segmentation was augmented as in the fourth experiment (Table \ref{tab:qantitative}, bottom row). 
The results show that correction of motion artifacts using motion correction network improves the performance (Table \ref{tab:qantitative}, second vs. third row). Moreover, results demonstrate that the performance of the segmentation network improves when the training data is augmented (Table \ref{tab:qantitative}, second row vs fourth row and third vs. bottom row).

\begin{table}[t]
\caption{Performance of brain tissue segmentation into eight tissue classes. The evaluation of segmentation was performed 1) on scans without motion artifacts (Motion Free) 2) on the same scans with synthesized motion using motion generation network (Motion Synthesized) 3) on the scans where synthesized motion were corrected using motion correction network (Motion Corrected). The segmentation network was retrained with motion augmented scans using motion generation network. The evaluation of segmentation was performed 4) on the scans with synthesized motion using motion generation network (Motion Augmented) 5) on the scans where synthesized motion were corrected (Motion Corrected and Augmented) }
\resizebox{\textwidth}{!}{
\npdecimalsign{.}
\nprounddigits{2}

\begin{tabular}
{|c|c|n{2}{2}n{2}{2}n{2}{2}n{2}{2}n{2}{2}n{2}{2}n{2}{2}n{2}{2}n{2}{2}|}

\hline
 &&{\enspace CB}& {mWM} & {\enspace BGT} & {vCSF} & {\enspace WM} & {\enspace BS} & {cGM} & {eCSF} & {Mean} \\ \hline

\multirow{3}{9em}{Motion Free}             & DC  & 0.90    & 0.53    & 0.89    & 0.84    & 0.94   & 0.84    & 0.67    & 0.83    & 0.80    \\
                           & HD & 44.92    & 32.97    & 39.06   & 23.08    & 17.25    & 42.57    & 18.47    & 8.60    & 28.36    \\
                           & MSD  & 0.36    & 1.85    & 0.56    & 0.36    & 0.20     & 0.56    & 0.21    & 0.23    & 0.54    \\
\hline                          

\multirow{3}{9em}{Motion Synthesized}         & DC  & 0.87    & 0.38    & 0.87    & 0.77    & 0.90    & 0.81     & 0.62    & 0.75    & 0.75    \\
						   & HD & 52.27    & 53.80     & 42.93    & 33.70    & 21.33    & 48.18     & 21.53    & 22.43    & 37.02    \\
                           & MSD  & 0.62    & 4.10    & 1.04    & 1.32    & 0.77    & 0.92    & 0.55    & 1.00    & 1.29     \\
\hline

\multirow{3}{9em}{Motion Corrected}           & DC  & 0.90    &  0.47    &  0.89    & 0.83     &  0.936    & 0.83    & \ubold{\enspace0.68}    &\ubold{\enspace0.85}    & 0.79     \\
                           & HD & 45.06    & 41.93    & 33.58   & 22.84    & 18.25    & 39.19     & 18.57    & 8.90    & 28.54     \\
                           & MSD  & 0.46     & 2.07    & 0.55    & 0.35    & 0.20    &\ubold{\enspace0.41}    & 0.207    &\ubold{\enspace0.16}    & 0.551    \\
\hline

\multirow{3}{9em}{Motion Augmented} 		   &DC & 0.88	&0.45	&0.88	&0.80 &	0.92 &	0.81 &	0.63	& 0.80 &	0.77\\
						   &HD&\ubold{40.19}	&\ubold{27.42}	&28.43	&19.27&	14.98   &\ubold{30.85}&	\ubold{15.03}&	11.79   &	23.49\\
						   &MSD&0.46	&1.84 &	0.61	&0.39	&0.27	&0.48	&0.27	&0.24	&0.57\\

\hline

\multirow{3}{15em}{Motion Corrected \& Augmented} 		   & DC  &\ubold{\enspace 0.91}	&\ubold{\enspace0.48}	&\ubold{\enspace0.89}&	\ubold{\enspace0.84}	&\ubold{\enspace0.94}&	\ubold{\enspace0.84}	&0.67	&0.84	&\ubold{\enspace0.80}\\
			   & HD &45.62	&34.52	&\ubold{26.83}	&\ubold{17.77}	&\ubold{14.40}	&35.93	&17.18	&\ubold{\enspace7.63}	&\ubold{24.99}\\
						   &MSD&\ubold{\enspace0.45}	&\ubold{\enspace1.89}	&\ubold{\enspace0.44}	&\ubold{\enspace0.29}	&\ubold{\enspace0.19}	&0.42	&\ubold{\enspace0.20}&	0.165	&\ubold{\enspace0.51}\\
\hline\end{tabular}

}

 \label{tab:qantitative}

\end{table}

\begin{figure}[t]
\includegraphics[width=\textwidth]{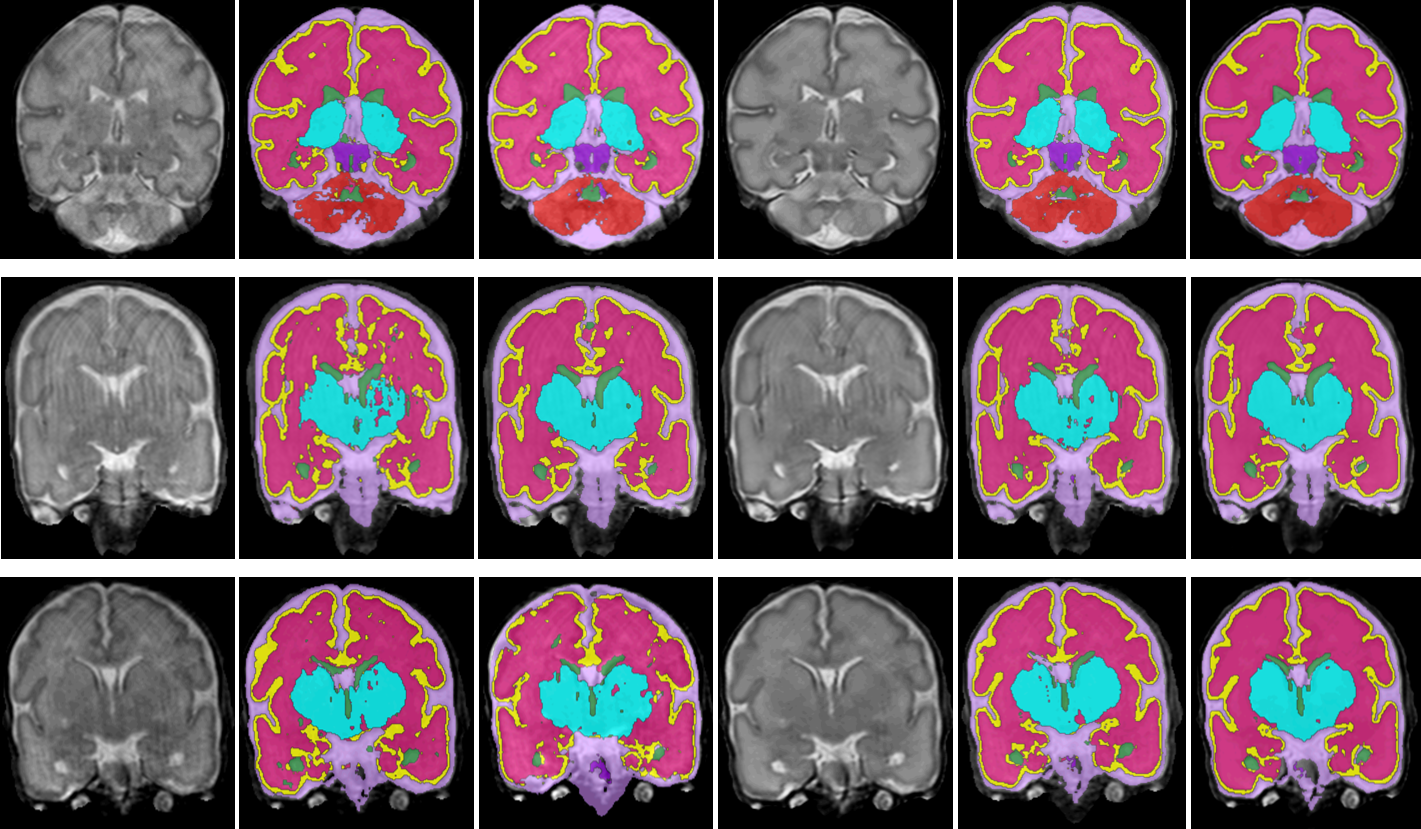}
\caption{ Examples of slices affected by motion artifacts and the corresponding tissue segmentation in neonatal MRI. 1st column: A motion affected slice; 2nd column: Automatic segmentation when the network was trained on slices without motion artifacts; 3rd column: Automatic segmentation, network trained on slices with augmented motion; 4th column: A motion corrected slice; 5th column: Automatic segmentation result on the corrected slice; 6th column: Automatic segmentation results on the corrected slice when the network was trained with data augmentation.}
    \label{fig:evaluation}
\end{figure}

To qualitatively evaluate the performance of the motion correction network, 40 scans affected by motion artifacts were corrected using motion correction network. Subsequently, the segmentation network trained with the proposed data augmentation was used to segment the corrected images. Qualitative scoring of the images and segmentations before and after motion correction was performed. The evaluation results show that the median image quality and quality of corresponding automatic segmentations were assigned grade 2 (poor) and 3 (moderate), respectively. After correction of motion artifacts, both improved to grades 3 and 4, respectively. Figure \ref{fig:evaluation} shows examples of images and corresponding segmentations before and after motion correction. This shows that the motion correction network reduces motion artifacts and hence, improves quality of the images and corresponding segmentations. Moreover, the figure shows that our proposed motion augmentation further improves automatic segmentations.

\begin{comment}

\begin{table}[t]
\setlength{\tabcolsep}{5pt}
\centering
\begin{tabular}{ccccc}
%\hline

&MA Image& MC Image& MA segmentation&MC Segmentation  \\
%\hline

%Q$_1$&1	&3	&2&	3\\
Median & 2	&3	& 3 & 4\\
%Q$_3$& 3	&4	&3,5&4\\

%&Q$_1$& Median& Q$_3$ \\
%\hline

%MA Image&1	&2	&3\\
 %MC Image & 3	&3	& 4 \\
 %MA segmentation& 2	&3	&3,5\\
%MC Segmentation & 3	&4	&4\\

%\hline
\end{tabular}

 \caption{The quality of scans and corresponding segmentation were graded using 5-point Likert scale. The median, and interquartile range is listed. Evaluation was performed in slices affected by motion artifacts (MA) and corresponding segmentations as well as in motion corrected (MC) slices with corresponding segmentations.}
\label{tab:score}
\end{table}

\end{comment}

\section{Discussion and conclusion}

We presented a method for correction of motion artifacts in reconstructed brain MR scans of preterm infants using a cycleGAN. We demonstrate that the proposed artifact correction generates images that are more suitable for (automatic) image segmentation. Additionally, we show that training the segmentation network with the proposed data augmentation further improves segmentation performance.

Unlike previous methods that performed motion correction in the frequency domain (k-space), the proposed method corrects motion artifacts in already reconstructed scans. Given that k-space data is typically not available after scans have been reconstructed and stored, the proposed method allows correction. %Given that k-space for motion in cases when raw data is no longer available.

To conclude, results demonstrate that correction of motion artifacts in reconstructed  neonatal brain MR scans is feasible. Moreover, results show that the proposed motion correction allows automatic brain tissue segmentation in scans affected by motion artifacts. This may improve clinical interpretability and extraction of quantitative markers in images with motion artifacts.

%In this study, the method was applied to neonatal MRI scans in infants imaged at 30 weeks PMA. However, this method can be applied to MR images visualizing different anatomy and acquired in different populations.

\bibliography{strings}

\end{document}